\documentclass[prl,twocolumn,showpacs,preprintnumbers,amsmath,amssymb,nofootinbib,superscriptaddress,10pt]{revtex4-2}
\usepackage{color}
\usepackage{graphicx}
\usepackage{slashed}
\usepackage{braket,amsmath,amssymb}
\usepackage{dsfont}
\usepackage{xcolor}

\usepackage{lettrine}
\usepackage{Zallman}

\usepackage{float}
\usepackage[colorlinks=true,linkcolor=blue,citecolor=blue,urlcolor=blue]{hyperref}
\usepackage{wasysym}
\usepackage[normalem]{ulem}
\usepackage{comment} %to use \iffalse \fi
\def\XXint#1#2#3{{\setbox0=\hbox{$#1{#2#3}{\int}$}
     \vcenter{\hbox{$#2#3$}}\kern-.5\wd0}}

\newcommand{\sgw}{\rho_{\mathrm{GW}}}

\newcommand{\cM}{{\cal M}}

\newcommand{\vq}{{\boldsymbol{q}}}
\newcommand{\vx}{{\boldsymbol{x}}}

\newcommand{\vK}{{\boldsymbol{K}}}

\newcommand{\vQ}{{\boldsymbol{Q}}}

\newcommand{\vk}{{\boldsymbol{k}}}

\newcommand{\bg}{\begin{align}}
\newcommand{\eeg}{\end{align}}
\newcommand{\be}{\begin{equation}}
\newcommand{\ee}{\end{equation}}
\newcommand{\ba}{\begin{eqnarray}}
\newcommand{\ea}{\end{eqnarray}}

\newcommand{\nn}{\nonumber}

\newcommand{\ve}{\varepsilon}

\newcommand{\p}{\omega_s}

\newcommand{\la}{\langle}
\newcommand{\ra}{\rangle}
\newcommand{\si}{\sigma}

\newcommand{\di}{d}
 
\newcommand{\ep}{\epsilon}

\newcommand{\om}{\omega}

\newcommand{\Dk}{\sigma_\omega}
\newcommand{\ks}{\omega_s}

\begin{document}

\title{Gravitational waves decay in vacuum}

\author{D. Blas\footnote{\href{mailto:dblas@ifae.es}{dblas@ifae.es}}}
 
\affiliation{\it Institut de F\'{i}sica d’Altes Energies (IFAE), The Barcelona Institute of Science and Technology, Campus UAB, 08193 Bellaterra (Barcelona), Spain \vskip 3pt
}
 
\affiliation{{\it Instituci\'{o} Catalana de Recerca i Estudis Avan\c{c}ats (ICREA),
Passeig Llu\'{i}s Companys 23, 08010 Barcelona, Spain}%\\ dblas@ifae.es 
\vskip 3pt}

\author{J.A. Oller\footnote{\href{mailto:oller@um.es}{oller@um.es}}}
\affiliation{{\it Departamento de F\'{\i}sica, Universidad de Murcia, E-30071 Murcia,  Spain}%\\ oller@um.es
}

\begin{abstract}
We show that gravitational waves (GW), treated as coherent graviton states, decay into photon pairs in vacuum.  The process, even if suppressed by $G^2$, is lifted by two effects combined: the expected factor of the graviton number squared, $N^2$, and the coherence of the wave. We perform the calculation describing both gravity and the photons as quantized fields, though we show that the effect admits a semiclassical description once the metric is solved to second order.
We estimate the resulting rates for compact binaries and a stochastic background, including the effect from stimulated decay to the cosmic microwave background (CMB). In theories with light degrees of freedom, an analogous decay into them is also possible, and more relevant for ultralight dark matter, as it can entail huge occupation numbers. We derive first constraints on cosmological GW sources by the corresponding injection of photons from CMB spectral distortions, extragalactic backgrounds, and light-nuclei photofission. In summary, the decay of GWs into photons offers a new (challenging) handle on the detection of GW sources, and represents a new mechanism to generate other particles across cosmic history.
\end{abstract}
 
\maketitle

%\tableofcontents
 
%%%%%%%%%%%%%%%%%%%%%%%%%%%%%%%%%%%%%%%%%%%%%%%%%%%%%%%%%%%%%%%%%%%%%%%%%%%%%%%%%%%%%%%%%%
%\section{Introduction}
%%%%%%%%%%%%%%%%%%%%%%%%%%%%%%%%%%%%%%%%%%%%%%%%%%%%%%%%%%%%%%%%%%%%%%%%%%%%%%%%%%%%%%%%%%
%{\bf Introduction.}  
\lettrine{T}{he} direct detection of gravitational waves (GWs) by the LIGO--Virgo--KAGRA collaboration \cite{LIGOScientific:2016aoc,LIGOScientific:2024elc}, the strong evidence from pulsar timing arrays \cite{NANOGrav:2023gor}, and ongoing efforts to probe a wide range of frequencies \cite{LISA:2022kgy,Aggarwal:2025noe,Blas:2026xol} have renewed interest in the precise prediction of GW signals. Most theoretical developments treat GWs as classical fields, with notable exceptions including GW generation in the early Universe \cite{Caprini:2018mtu}, probes of the quantum nature of gravity \cite{Dyson:2013hbl,Parikh:2020kfh,Tobar:2023ksi,Carney:2023nzz,Kanno:2025how,Berezhiani:2024boz}, or in studies of the infrared properties of quantum processes of matter or gravitation, see e.g. \cite{Flauger:2019cam,Ai:2025xla}, as well as investigations of infrared-enhanced quantum collective effects in graviton systems \cite{Sawyer:2019plp}. Given the large occupation numbers of the states emitted by most observable sources \cite{Moffat:2024gkj,MacKay:2024sgw}, it is arguable that a classical description is generically sufficient.

At a more fundamental level, GWs are quantum objects corresponding to coherent states of gravitons---the minimum-uncertainty states that most closely resemble classical waves---with dynamics governed by general relativity interpreted as an effective field theory (EFT) below the Planck scale, $M_P \sim 10^{19}\,\mathrm{GeV}$ \cite{Donoghue:2017pgk,Burgess:2003jk}, interacting with the Standard Model. As a result, the evolution of the GW has to be studied with these coherent states as initial states, which, as we will see, implies their decay into light particles, photons in particular. The generic formalism to treat this and other dynamical effects is quantum for \emph{both gravitation and electromagnetism}, and only on certain occasions a semiclassical or classical description is appropriate. Coming back to our process of interest, the quadratic coupling of photons forbids it at the classical level from pure vacuum (this is different if one considers decays in a medium). Regarding semiclassical calculations, no particle production occurs for an individual plane-wave GW  
\cite{Schwinger:1951nm,Gibbons:1975jb,Deser:1975ffb,Dunne:2004nc,Khusnutdinov:2025len,Garriga:1990dp} (see \cite{Dorca:1993sv,Redi:2026nzi} for other limits where the process exists, and GWs are involved). 

In the rest of this \textit{letter}, we will go beyond this limit and study the GW decay from a quantum field theory perspective.  This will show two interesting features: first, that the process exists and grows strongly with frequency. We also show that it admits a semiclassical description, which requires taking into account the nonlinearity of General Relativity; second, that it is enhanced by the number of gravitons as $N^2$ in a peculiar way because of coherence, which is absolutely necessary to even think about beating the smallness of the proportionality factor $M_P^{-2}$ in the rates.  %arriving from the smallness of $M_P^{-1}$. 
 We will derive this effect and study some first ideas on its phenomenology. We also discuss the possible decay to other light bosons, as may happen in the presence of ultralight dark matter (ULDM). We use natural units $c\!=\!\hbar\!=\!1$ (unless the opposite is stated).

\vspace{.1cm}
%%%%%%%%%%%%%%%%%%%%%%%%%%%%%%%%%%%%%%%%%%%%%%%%%%%%%%%%%%%%%%%%%%%%%%%%%%%%%%%%%%%%%%%%%%
\noindent {\bf Coherent states of gravitons.} The $S$-matrix corresponding to a classical source with energy-momentum $T^{\mu\nu}_s$ coupled linearly to the EFT of gravity is  given by
\begin{equation} \label{250812.1}
\hat S=\lim_{\substack{t_f\to+\infty \\ t_i\to-\infty}} T \, \exp\left(i(8\pi G)^{1/2}\int_{t_i}^{t_f} \!\!dt\int d^3x\,   \hat h_{\mu\nu} T_s^{\mu\nu}\right),
\end{equation}
where  $G$ is the Newton constant and $\hat h_{\mu\nu}(x)$ is the field-operator %$h_{\mu\nu}$ 
of the metric perturbations around the Minkowski metric $\eta_{\mu\nu}$ \cite{Donoghue:2017pgk} (we follow the same conventions as in \cite{Blas:2020dyg}),
\begin{align}
\label{250715.2}
\hat h_{\mu\nu}(x)=\sum_\lambda\int\frac{\di^3k}{(2\pi)^3 2 k}\ve^\lambda_{\mu\nu}(\vk)\hat a_\lambda(\vk)e^{-ik\cdot x}+h.c.
\end{align}
Here, $\ve^\lambda_{\mu\nu}(\vk)$ are the polarization tensors and $\hat a_\lambda(\vk)$ the annihilation operators of helicity $\lambda=\pm 2$. 
From Eq.~\eqref{250812.1}, the final state after the source is active is $|f\rangle= \hat S|0\rangle$, a coherent state expressed in Fourier space as\footnote{Some expressions differ from the quantum-optics conventions because our operators satisfy $[\hat a(\vk),\hat a^\dagger(\vk')]=(2\pi)^32k\delta^{(3)}(\vk-\vk')$.}
\begin{align} \label{250812.8}
  |f\ra&=e^{{\sum_\lambda\int \frac{d^3k}{(2\pi)^3 2k} \left( f^s_\lambda ({\vk})\hat
  a_\lambda(\vk)^\dagger-\text{h.c.}\right)
  }}|0\ra\,\equiv \hat D(f_\lambda)|0\ra,
\end{align}
with $f^s_\lambda({\vk})=i(8 \pi G)^{1/2}\,\ve^\lambda_{\mu\nu}(\vk)^* T_s^{\mu\nu}(k,\vk)\,$, with $k=|\vk|$ (on-shell Fourier transform of the source). The state  $|f\ra$  is  an eigenstate of the annihilation operators,
\begin{align}
 \label{250702.2}
 \hat a_\lambda(\vk)|f\rangle=f^s_\lambda(\vk)|f\rangle\,.
\end{align}
Since $\hat D(f_\lambda)$ satisfies $
   \hat  D^{\dagger}(f_\lambda)\hat  a_\lambda(\vk) \hat D(f_\lambda)=\hat a_\lambda(\vk)+f^s_\lambda(\vk) \hat{\mathds{1}}$ \cite{Ilderton:2017xbj}, the dynamics is equivalent to substituting $\hat h_{\mu\nu}$ by $\hat h_{\mu\nu}+h^{c}_{\mu\nu} \hat{\mathds{1}}$ in the quantum theory, where 
\begin{equation}
\label{eq:classic}
 \hspace{-.1 cm}   h^c_{\mu\nu}=\langle {f}| \hat h_{\mu\nu}|  {f}\rangle=\sqrt{8\pi G}\! \int \!\di^4y\, \Delta_{\rm rad}(x-y)_{\mu\nu\alpha\beta} T_s^{\alpha\beta},
\end{equation} 
is the classical radiated field by the source, and $ \Delta_{\rm rad}(x-y)_{\mu\nu\alpha\beta}$ is the Pauli-Jordan function \cite{Ilderton:2017xbj,Itzykson:1980rh}.

Assuming a localized source with frequency centered around $\ks$ with width $\Dk\ll \ks$ (we suppress the polarization label from now on)
$
f^s\propto 
\exp\!\left(-\frac{(k-\ks)^2}{2\Dk^2}\right),
$ 
and far from the source,  the classical field of Eq.~\eqref{eq:classic} tends asymptotically to the outgoing spherical wave packet,  
$h^c_{\mu\nu}(r-t)\propto \frac{\cos(\ks(r-t))}{r}\,\exp\!\left[-(r-t)^2\Dk^2/2\right]$. 
This waveform corresponds to a burst of duration $\delta t\sim 1/\sigma_\omega$ and characteristic frequency $\omega_s$, approaching continuous emission for large $\delta t$. We will comment on the validity of these approximations for each source we shall consider. For convenience, we approximate the previous profile by a narrow-band top-hat distribution, $f^s(k)=\Theta$ for $k\in[\ks-\tfrac{\Dk}{2},\,\ks+\tfrac{\Dk}{2}]$ and zero otherwise, with $\Theta=\sqrt{2\pi^2N/\om_s\si_\om}$ fixed by the average graviton number $N$.

\begin{figure}
    \begin{center}    \includegraphics[width=0.7\columnwidth]{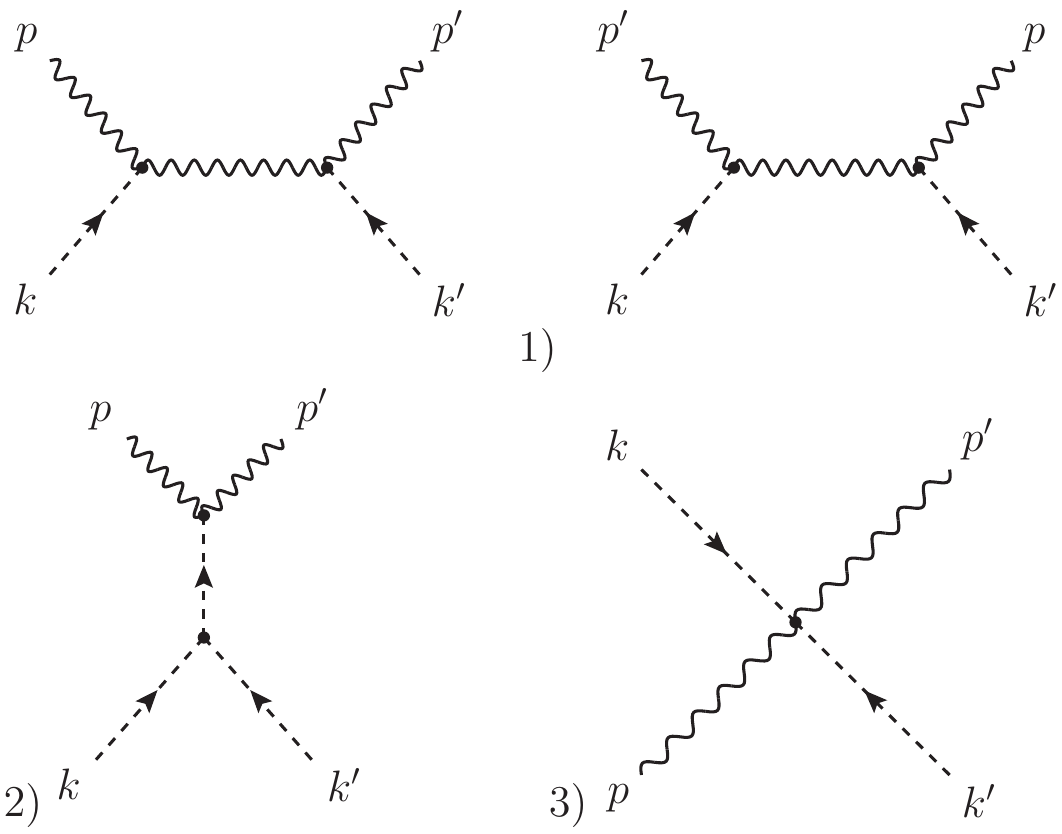}
\caption{{\small Feynman diagrams for the scattering amplitude $g(\vk)+g(\vk')\to\gamma(\boldsymbol{p})+\gamma(\boldsymbol{p}')$, first calculated in Ref.~\cite{Skobelev:1975gpi}.}
\label{fig.150715.1}}
    \end{center}
    \vspace{-0.5cm}
\end{figure}

\vspace{.1cm}
%%%%%%%%%%%%%%%%%%%%%%%%%%%%%%%%%%%%%%%%%%%%%%%%%%%%%%%%%%%%%%%%%%%%%%%%%%%%%%%%%%%%%%%%%%%%%%%
\noindent {\bf Graviton fusion in coherent states and radiation of photons.} The leading-order amplitude in perturbative gravity for two-graviton fusion into photons, $
g(\vk_1,\lambda_1) +g(\vk_2,\lambda_2)\to \gamma(\boldsymbol{q}_1,\mu_1)+\gamma(\boldsymbol{q}_2,\mu_2)$ (momenta and helicities in brackets), was computed by Skobelev \cite{Skobelev:1975gpi} (see also \cite{Grisaru:1975bx,Bjerrum-Bohr:2014lea}). The corresponding Feynman diagrams are shown in Fig.~\ref{fig.150715.1}, and the fusion amplitude $A_{\lambda_1\lambda_2\mu_1\mu_2}(k_1,k_2,p_1,p_2)$ reads 
\begin{align}
\label{250530.3}
A_{++++}\!&=\!A_{----}=8\pi G\frac{{\rm t}^2}{{\rm s}}\,,\\
A_{++--}\!&=\!A_{--++}=8 \pi G\frac{{\rm u}^2}{{\rm s}}\,,\nn
    \end{align}  
where ${\rm s}=(k_1+k_2)^2$, ${\rm t}=(k_1-q_1)^2$, and ${\rm u}=(k_1-q_2)^2$ are the usual Mandelstam variables.  
The initial gravitons can also scatter into gravitons outside the coherent state, with rates of the same order (up to a different Bose enhancement); we proceed with the diagrams of Fig.~\ref{fig.150715.1}, which introduce the phenomenology associated with photon injection and the corresponding depletion of gravitons.

\vspace{.1cm}
%___________________________________________________________________________
\noindent From Eq.~\eqref{250702.2}, the annihilation operators in the Fourier decomposition of the graviton field act on $|f\ra$ by multiplication with the profile $f^s(k)$.\footnote{When sandwiched between the coherent states, other contributions involving $a^\dagger a$  are suppressed because of energy conservation in the narrow-band limit $\Dk/\p\ll 1$.} In addition, there are {\it standard} graviton propagators as in diagram 2) of Fig.~\ref{fig.150715.1}. Contributions from an intermediate GW coherent state vanish, as they require an on-shell intermediate graviton (belonging to the GW), for which $s=0$, which corresponds to a vanishing scattering amplitude (see Supplemental Material (SM), Sec.~\ref{app.260319.1} for more details). 
It hence follows from Eq.~\eqref{250530.3} that the fusion amplitude, assuming an unpolarized graviton state,  
can be expressed as 
\begin{align} \label{250715.7}
A_{\mu_1\mu_2}(\vq_1,\vq_2)&\!=\!\frac{4\pi G}{(2\pi)^6}\!\int\frac{{\di}^3k_1}{2k_1}\!\int\frac{{\di}^3k_2}{2k_2}f^s(k_1)f^s(k_2)
\\
& \times \frac{{\rm s}^2-2{\rm t}{\rm u}}{{\rm s}}(2\pi)^4\delta(q_1+q_2-k_1-k_2)\delta_{\mu_1\mu_2}.\nn
\end{align}
This expression is based on the use of the Feynman propagator in diagram 2) of Fig.~\ref{fig.150715.1}. From the relation between the Feynman and retarded propagator $G_F(k)=G_{\rm ret}(k)-2\pi i\,\theta(-k^0)\,\delta(k^2)$ \cite{Itzykson:1980rh}. As happened before, this expression contains a term (the second one) which vanishes when convoluted with the external coherent states. As a result, the calculation we will perform is equivalent to that of considering the creation of photons in a background $h_c$, once considered at second order (including in particular the induced metric sourced by the energy-momentum tensor of the GWs, cf. Fig.~\ref{fig.150715.1}). To the best of our knowledge, this also supersedes previous semiclassical calculations. Regarding other processes, $g\to g\,\gamma\,\gamma$ has a vanishing
rate, and $\emptyset\to g\,g\,\gamma\,\gamma$ violates energy-momentum
conservation. The process $g\to\gamma\gamma$ is forbidden by the graviton
being massless. Multiphoton production carries higher powers of $G$ and
suffers a large phase-space suppression for massless multiparticle
states \cite{Blas:2020dyg,Escudero-Pedrosa:2020rwb}. 
 
The decay probability $P$ follows from integrating the square of Eq.~\eqref{250715.7} over all photon momenta and summing  their helicities,  with a factor $1/2$ for photon indistinguishability, giving (see SM~\ref{sec.260611.1}) 
\begin{align} \label{250717.10}
\frac{dP}{dt}&=\frac{3G^2N^2\ks^3\Dk^2}{25\pi}\,.
\end{align}

Each reaction removes two gravitons and generates two photons. As a result, the number of photons radiated is $N_\gamma(t)=N_0-N(t)$, with $N_0$ the initial graviton number and, at leading order in $G$ (see SM~\ref{sec.260611.1}), 
\begin{align} \label{250717.11}
N_\gamma(t)\approx G^2N_0^2\frac{6\ks^3\Dk^2 t}{25\pi}.
\end{align}
From this equation, the characteristic reaction time is
\begin{align} \label{250717.11b}
\tau&=
\frac{25\pi}{6 G^2E_{\text{rel}}^2\ks\sigma_\omega^2}\,,
\end{align}
with $E_{\text{rel}}=N_0\,\omega_s$ the total energy in the wave front of width $\sigma_\omega$. Equation~\eqref{250717.10} is a probability per unit time for $\tau \ll\delta t \sim 1/\sigma_\om$, with Poissonian uncertainty $\sqrt{\sigma_\omega\, \tau}$. 

The radiated electromagnetic power $W_\gamma$ then reads
\begin{align} \label{250717.13b} 
W_\gamma=\!\ks\frac{d N_\gamma}{d t}\approx\! \frac{6 G^2 }{25\pi} E_{\text{rel}}^2 \p^2\Dk^2\!=\!\frac{6 G^2  \hbar}{25\pi c^{10}} E_{\text{rel}}^2 \omega^2\delta_\omega^2  \,.
\end{align}
Restoring SI units, $\ks=\hbar \omega$, $\sigma_\omega=\hbar\delta_\omega$. Being linear in $\hbar$ means that the quantumness of this number comes from the energy of photons, confirming the semiclassical origin for the gravitational field. Defining the gravitational radius $r_m=2GE_{\text{rel}}/c^4$, and the wavefront width $\ell=c\,\delta t$, \ we can re-express $W_\gamma$  as
\begin{align}
W_\gamma= \frac{3}{50\pi}\left(\frac{r_m}{\ell}\right)^2\hbar\,\omega^2.
\end{align}

For the burst-like configuration, the strain $h_r$ at distance $r$ satisfies $
\frac{E_{\rm rel}\sigma_\omega}{4\pi r^2}
\approx\frac{\omega_s^2 h_r^2}{32\pi G}$; and is reduced as the GW decays, depleting at time 
$\tau_h=\tau E_{\rm rel}/\omega_s$ (cf.  Eq.~\eqref{250717.11b}). 

When the GW propagates through an isotropic unpolarized photon bath of occupation number $f_\gamma(q)$ (e.g.\ the CMB),
the net balance of stimulated emission and absorption replaces\footnote{Since stimulated emission/absorption effects are relevant only for $\ks \lesssim k_BT$, we can also take that $\sigma_\omega/k_B T\ll 1$. As the photon energy in the laboratory frame depends on the direction of momentum, this effect is considered in the average sense.} $W_\gamma\to W_\gamma (1+2f_\gamma(\p))$ in all the previous formulae. In the presence of a background, one can see that all the $\hbar$ disappear, and our calculation corresponds to a classical process.

\begin{figure*}
\centering
\includegraphics[width=0.46\textwidth]{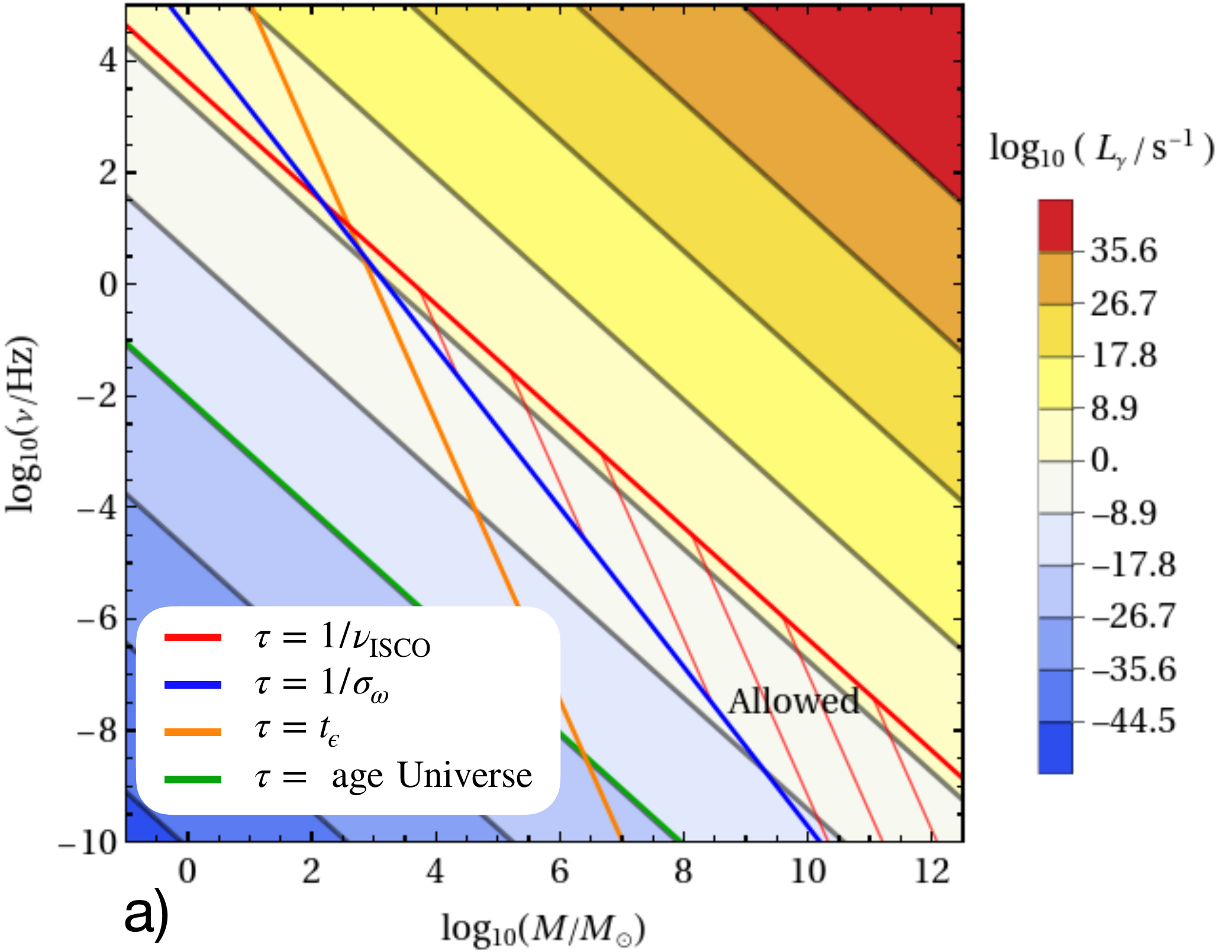}\hfill
\includegraphics[width=0.46\textwidth]{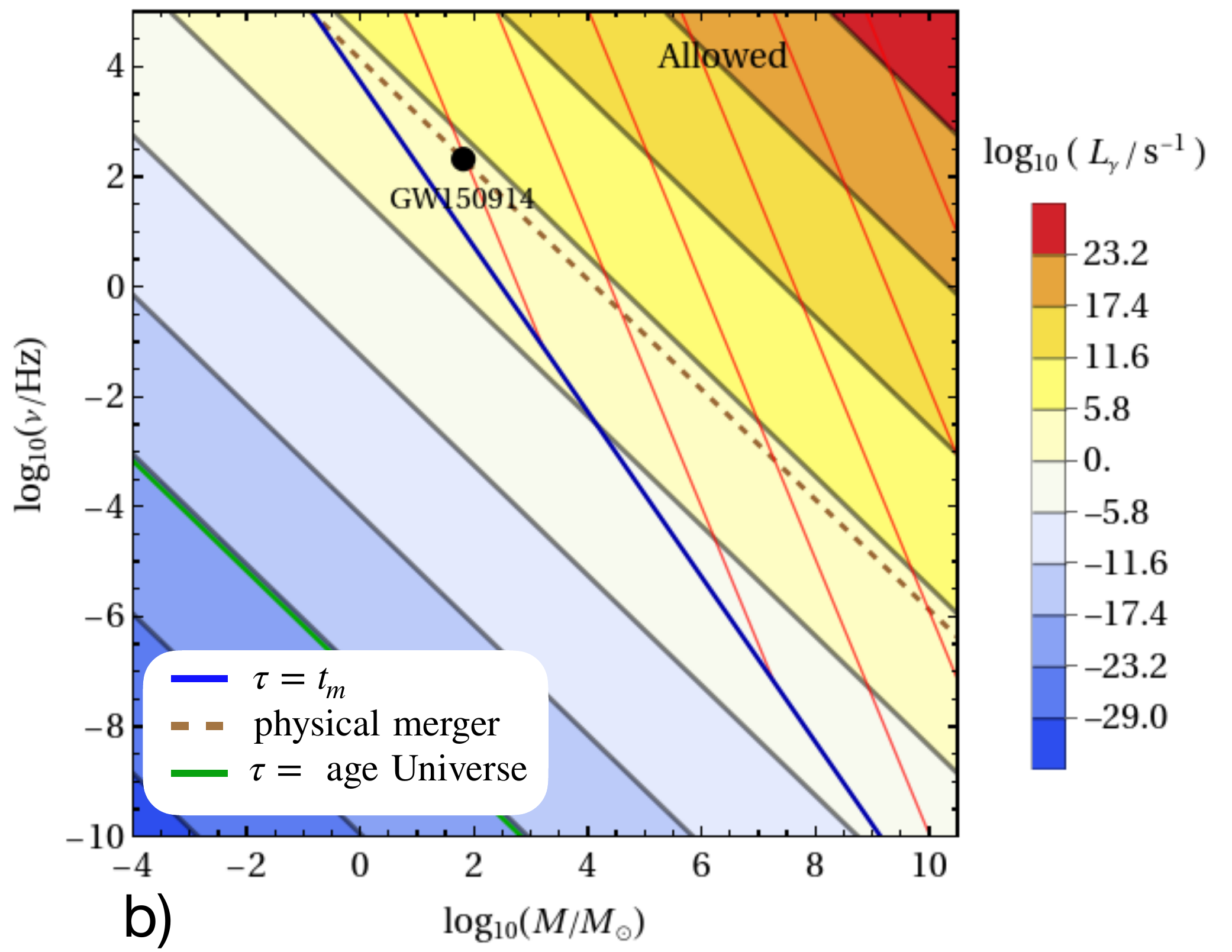}
\caption{{\small   
Photon luminosity $L_\gamma$ radiated by the GW  emitted by a compact binary, shown as log--log contour plots in the plane of total mass $M$ and GW frequency $\nu$, for the inspiral [a), left] and merger [b), right]
phases. The hatched areas indicate the regions allowed by the constraints discussed in the text. Both cases include the possible enhancement from the average occupation numbers of photons today. See the text for further details.}\label{fig.260611.1}}
\end{figure*}

%%%%%%%%%%%%%%%%%%%%%%%%%%%%%%%%%%%%%%%%%%%%%%%%%%%%%%%%%%%%%%%%%%%%%%%%%%%%%%%%%%%%%%%%%%%%%%%
\vspace{0.1cm}
\noindent{\bf Application to binary systems.}
We now apply the conversion rate of Eq.~\eqref{250717.10} to the GWs emitted by binary systems. 
We treat only bound binaries, since unbounded orbits yield results not far from the merger case, see e.g. \cite{Caldarola:2023ipo}.  For the \textit{inspiral phase}, we consider the quasi-circular inspiral of a
binary of chirp mass $\cM_c$. Using the standard quadrupole formulas
for the GW luminosity $L_{\text{GW}}$ and the frequency drift
$d\p/dt$ \cite{Maggiore:2007ulw}, the time during which the frequency grows from $\p$
to $\p(1+\ep)$ is $t_\ep=\frac{5\,\ep}{12\left(2(G\cM_c)^5\p^8\right)^{1/3}}$,
and the gravitational energy released in this interval is
$E_{\text{rel}}=t_\ep L_{\text{GW}}$. From $E_{\text{rel}}$ and $\p$
we evaluate the reaction time $\tau$ of Eq.~\eqref{250717.11b} and
impose $\tau \ll \min\!\left(1/\ep\p,\,t_\ep\right)$,
so that photon emission takes place well within the coherence time
$1/(\ep\p)$ of the wave train, while  the binary sweeps the
corresponding frequency interval. Introducing the number of cycles $n=\p t_\epsilon/2\pi$, if $t_\epsilon<1/(\ep \p)$ then $\Dk\sim 1/t_\epsilon$ and $\Dk/\p\sim 1/(2n\pi)$. 
Dividing the GW into fractional bandwidth bins
 $\ep$ yields a lower bound on the emissivity, as any
residual coherence between bins is neglected. However, four terms are involved by squaring the modulus of Eq.~\eqref{250715.7} to calculate the probability, and they increase the decoherence as $\ep$ grows.
We further require the frequency
$\nu=\p/2\pi$ to lie below the ISCO value, $\nu_{\text{ISCO}}$, which delimits the
inspiral regime.  
Figure~\ref{fig.260611.1}a) shows a log-log contour plot of the
photon-number luminosity $L_\gamma=W_\gamma/\p$ over the total mass $M$ and
frequency $\nu$ for an equal-mass binary ($M=2^{6/5}\cM_c$, $\ep=0.1$).
The hatched allowed region is bounded from above by $\nu_{\text{ISCO}}$ frequency
(red line) and from below by the condition of whichever
timescale is shorter: the blue line [$\tau=1/(\ep\p)$] at low
frequencies or the orange line ($\tau=t_\ep$); $\tau$ then never
exceeds the age of the universe (green line) in the region shown. Interestingly, a
genuine window of applicability of our formalism therefore exists. However, the resulting 
 $L_\gamma$ is modest, and the frequencies are below kHz.\footnote{The flux can be estimated by interpreting this luminosity as the number 
of photons produced per unit time at the sphere of radius $\tau$. Note that if the frequencies are below kHz, they are immediately absorbed by the interstellar medium~\cite{Ocker:2026mta}.}

For the \textit{merger phase}, at fixed mass ratio and spins, the characteristic
frequency $\p$, the released energy, and the duration $t_m$ of the phase, of around one period,  scale as \cite{Buonanno:2006ui,Flanagan:1997sx}: $\nu %= \omega_s/2\pi
\approx 150\,{\rm Hz}\times (65\,M_\odot/M)$, $E_{\rm rel} \approx 2\,M_\odot c^2 \times (M/65\,M_\odot)$ and
$t_m \approx 5\,{\rm ms} \times (M/65\,M_\odot)$, where we anchor the total mass to GW150914
(recall the universality properties of mergers of black holes). 
We again obtain $\tau$ from
Eq.~\eqref{250717.11b} and require $\tau\ll t_m$ and $\tau$ smaller
than the age of the universe.\footnote{For $\ep=0.1$  the period $1/\nu$ and the coherence time $1/\ep\, \p$ roughly coincide. In the ringdown phase of GW150914 $\ep\approx 0.15$ \cite{Carullo:2019flw}.}  
The resulting $L_\gamma$ is shown in Fig.~\ref{fig.260611.1}b),
with the hatched allowed region bounded by the blue ($\tau=t_m$) and
green ($\tau=$ age of the universe) lines. Physical mergers lie on the
dashed brown line, the nominal merger frequency rescaled from
GW150914 (black dot). At the low frequencies of the allowed region,
the CMB occupation number conspires to make the number luminosity of physical mergers scale invariant $\sim M^0$, as clearly seen along the brown line of the figure. Along the nominal frequency
$L_\gamma$ is larger than in panel a), though still modest,
$\sim 10^6~\text{s}^{-1}$. In the leftmost corner of the allowed region in  Fig.~\ref{fig.260611.1}b), frequencies are above keV and may propagate towards our detectors.

It is instructive to compare $W_\gamma$ with the Hawking luminosity $W_H$ of a black hole of mass corresponding to the total mass $M$,
\begin{align}\label{260630.1}
\frac{W_H}{W_\gamma}\!=\!\frac{5}{4608\pi^2}\left(\frac{\lambda\, \ell}{r_sr_m}\right)^2\!\!\approx  10^{-4}\left(\frac{\lambda\, \ell}{r_sr_m}\right)^2\!\!,
\end{align}
with $r_s=2GM$, the Schwarzschild radius. $W_H/W_\gamma$ is proportional to the square of the ratio between two characteristic areas: a wavefront area $\lambda \ell$ and a gravitational area $r_s r_m$.  For GW150914 one finds $W_H/W_\gamma\!\sim\!10^2$--\,$10^3$, scaling with the binary masses in the merger phase like $M^0$.
A key difference between both sorts of radiation is that the one from the decay of the wavefront happens in an extended region.

%%%%%%%%%%%%%%%%%%%%%%%%%%%%%%%%%%%%%%%%%%%%%%%%%%%%%%%%%%%%%%%%%%%%%%%%%%%%%%%%%%%%%%%%%%%%%%%
\vspace{0.1cm}
\noindent{\bf Stochastic gravitational-wave background}. We now consider the $gg\to\gamma\gamma$ amplitudes for a stochastic gravitational-wave background (SGWB). 
For an occupation number $N_k\gg1$, squaring Skobelev's amplitudes, averaging the SGWB distribution [SM Eq.~\eqref{260401.1}], and integrating over photon momenta, with their polarizations summed, yield the decay rate per volume 
\begin{align} 
\label{260401.3}
\frac{d P}{dtdV} =
    \frac{8\pi G^2}{15}\,
      \sgw^2\,,
\end{align}
with $\rho_{\text{GW}}$ the GW energy density. The number emissivity is absolutely negligible, $2 dP/dtdV\approx 10^{-110}$~m$^{-3}$s$^{-1}$, assuming the $N_{\rm eff}$ bounds on the energy density of GWs~\cite{Caprini:2018mtu,Planck2018VI}. 
Comparing this rate with that of the pure-coherent state, Eq.~\eqref{250717.10}, the difference amounts to the 
replacement $V\omega_s\sigma_\omega^2\to 40\pi^2/9$: for a coherent state the 
effective volume is reduced to $5\lambda_s^3 /(9\pi\epsilon^2)$,  roughly a wavelength volume for the values of $\epsilon$ we consider, dramatically enhancing the 
effective density relative to the incoherent case.

From the relation $\frac{d\sgw}{d\ln\! k}= \frac{k^4}{\pi^2}\!\left(N_k+\frac{1}{2}\right)$, this rate of decay can be translated into a close equation for $\rho_{\rm GW}$. In terms of the variable $\rho^{\text{co}}(z)\equiv \rho_{\rm GW}/(1+z)^4$, which includes the appropriate redshift evolution, one finds (cf. Eq.~\eqref{260401.4})
\begin{equation}
    \frac{d\rho^{\text{co}}(z)}{dk}=e^{-k\,\Phi(z)}\frac{d\rho^{\text{co}}(0)}{dk},
\end{equation}
with $\Phi(z)\! =\! -\!\int_0^z \!dz'\, r(z')\,\rho(z')$ and $r(z)=16\pi G^2(1+z)^4/15H(z)$.
This equation can be treated perturbatively. Up to the GUT scale ($z\sim 10^{28}$), $|\Phi|\approx 10^{-25}\,\text{Hz}^{-1}$ for $\rho(0)_{\rm GW}$ at its largest value compatible with bounds on $\Delta N_{\text{eff}}$ \cite{Planck2018VI}: the depletion seems negligible for all range of frequencies.

%%%%%%%%%%%%%%%%%%%%%%%%%%%%%%%%%%%%%%%%%%%%%%%%%%%%%%%%%%%%%%%%%%%%%%%%%%%%%%%%%%%%%%%%%%%%%%%
\vspace{0.1cm}
{\bf Graviton fusion in coherent states beyond the standard model: ULDM example.} The previous sections show that the
rates to generate particles out of gravitational waves seem too small to have a clear significant effect. This may change in models deviating from the standard scenario. Interesting possibilities include theories with stronger gravitational interactions at small scales \cite{Csaki:2004ay,Dvali:2007wp}, with extra attractive forces operating at small scales to unlock $\omega_s$ from $M$, or in decays  $gg\leftrightarrow\phi\phi$ to particles $\phi$ with very large occupation numbers. 
Ultralight dark-matter (ULDM) candidates of masses $m_{\rm DM}\ll 1 \,$eV fit into the last point \cite{Hu:2000ke,Hui:2016ltb}: 
at the cosmological mean DM energy density
\begin{equation} \label{260505.1}
  N_\phi%
  \simeq 6\times 10^{87}\left(\frac{10^{-22}\,\rm eV}{m_\text{DM}}\right)^4 \left(\frac{10^{-3}}{v}\right)^3\,
  \frac{{\Omega_\text{DM}}}{0.26} x_{\rm MW},
\end{equation}
where $v$ represents the velocity dispersion, $\Omega_{\rm DM}$ its average relative energy density and  $x_{\rm MW}$ is the boost one gets in a DM halo; for the Milky Way halo $x_{\rm MW}\sim 10^5$~\cite{Read:2014qva}. The elementary helicity amplitudes for  $gg\leftrightarrow\phi\phi$ follow, e.g., from \cite{Gross:1968in} crossed to the $t$-channel. We model the ULDM background as a stochastic background following \cite{Cheong:2024ose,Foster:2020fln}. After completing the calculation (cf. SM~\ref{sec.smscalars}), 
for $\omega_s\gg m_{\rm DM}$ the decay happens into states basically unoccupied and  $dP/dt= 7\,G^{2} N_0^{2}\p^{3}\Dk^{2}/450\pi$.  The resulting reaction time is $\tau_\phi=(54/7)\,\tau$ [$\tau$ from Eq.~\eqref{250717.11b}]. In the regime  where the decay happens into states of high occupation number,  $m_\text{DM}\leq\p\lesssim  m_\text{DM}(1+v^{2}/2)$, $N_\phi$ can enhance the rate enormously, 
\begin{equation} \label{260505.2}
  \frac{dP}{dt}\simeq 
  \frac{G^{2}m_\text{DM}^{5} N_0^{2} v^{6}}{128}\,N_\phi,
\end{equation}
with  $\Dk
\approx m_\text{DM} v^2/2$.
Using Eq.~\eqref{260505.1}, the converted net power is $\displaystyle{W_\phi= \frac{9\pi\, G\, m_\text{DM}^{2} v^{3} N_0^{2}}{256}H_0^{2}\,\Omega_\text{DM}}$. For a binary merger of total mass $M$, since for ISCO frequencies $m_{\rm DM}\sim\omega_s\sim 8800\pi(M_\odot/M)$~\cite{Maggiore:2007ulw}, $N_0\propto m_{\rm DM}^{-2}$ and $W_\phi\propto m_{\rm DM}^{-2}$, favoring lighter ULDM.  Hence, 
\begin{equation}
  W_\phi\sim 10^{14}\,x_{\rm MW} \left(\frac{10^{-21}\,\rm eV}{m_{\rm DM}}\right)^2\left(\frac{v}{10^{-3}}\right)^3\, {\rm W},
\end{equation}
corresponding to rates of emission of $\sim 10^{54}\,$s$^{-1}$ for $m=10^{-21}\,$eV.

%%%%%%%%%%%%%%%%%%%%%%%%%%%%%%%%%%%%%%%%%%%%%%%%%%%%%%%%%%%%%%%%%%%%%%%%%%%%%%%
\vspace{.1cm} \noindent {\bf Phenomenological bounds on GW density rates.} 
%%%%%%%%%%%%%%%%%%%%%%%%%%%%%%%%%%%%%%%%%%%%%%%%%%%%%%%%%%%%%%%%%%%%%%%%    
We now discuss the implications of our previous results for unspecified sources.  We treat $E_{\text{rel}}$ and $\p$ as independent parameters and study the observable impact of the energy deposited by photons radiated from a homogeneous population of recurrent GW events. 
The sources will have comoving density $\tilde{n}$ and a rate of emission $\varkappa$ from time $t_a$ onward.

Summing the energy radiated by all sources causally connected with the observation point, in the limit  $1/\varkappa(t-t_a)\ll 1$ and $H(z)/\varkappa\ll 1$, the total deposited energy density reads (see SM~\ref{sec.smbounds} for the derivation),
\begin{align}\label{260122.3}
\frac{\Delta\rho_\gamma(z,\p)}{(1+z)^4}&\approx\!\,\frac{W_\gamma\Gamma }{H_0^2}
F(z,\omega)\,,
\end{align}
where $\Gamma=\tilde{n}\varkappa$, and  $F(z,\omega)$ is a dimensionless function, see Eq.~\eqref{eq:Fzomega}. 
The injection of radiation into the Universe is a well-explored direction to study standard and new physics \cite{Zeldovich:1969ff,Chluba:2018cww,Chluba:2019nxa,Masso:1997ru}. The conclusions depend on the redshift and band of injection. Far from aiming at a complete study, we now discuss a few possibilities to constrain $\sqrt{\Gamma}\, \ep E_{\text{rel}}/M_\odot$ as a function of the frequency of the GWs.

\begin{figure}[ht]
\begin{center}
\includegraphics[width=\columnwidth]{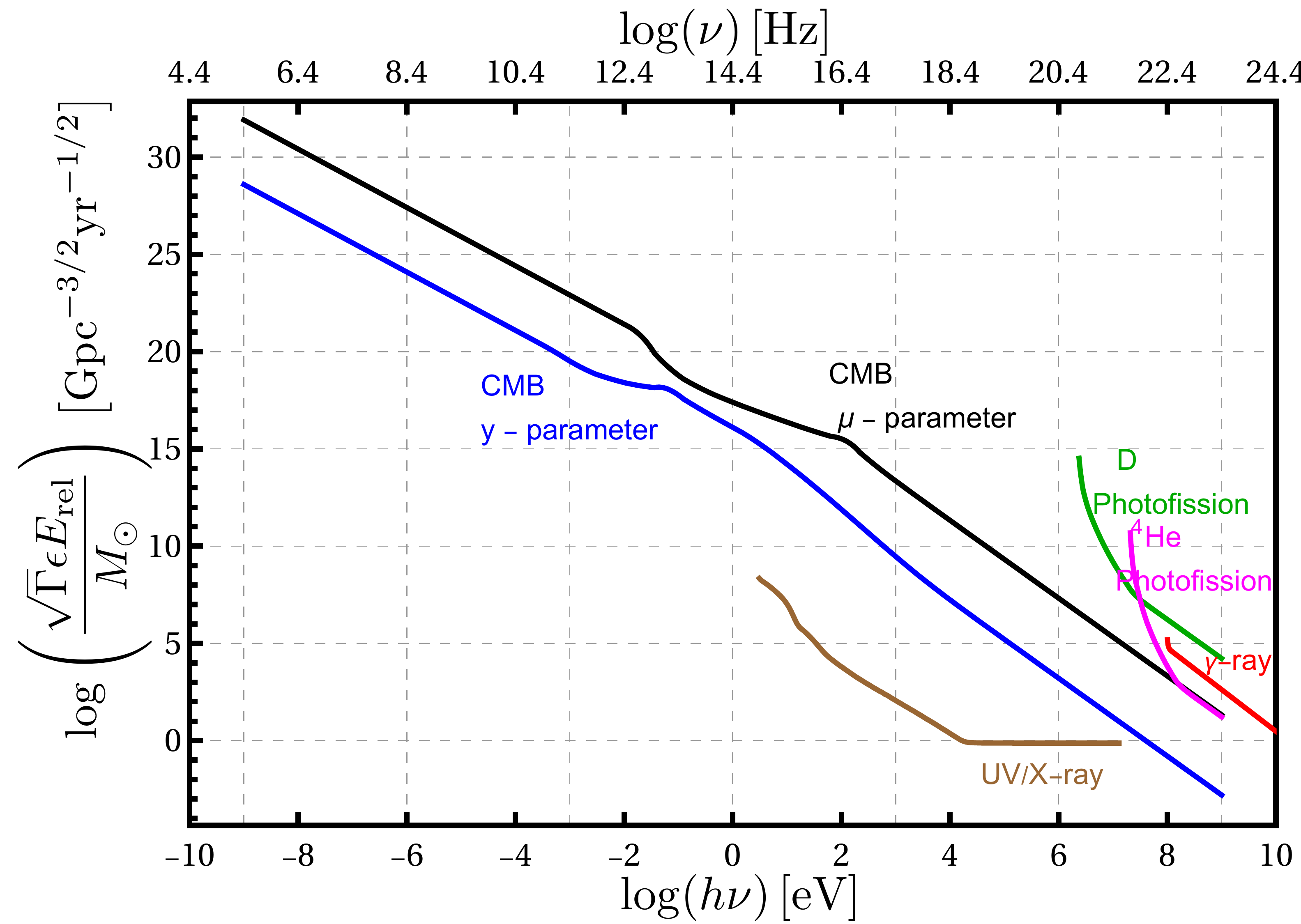}
\end{center}
\caption{{\small 
Log--log plot showing the upper bounds on $\sqrt{\Gamma}\,\epsilon E_{\text{rel}}/M_\odot$~[Gpc$^{-3/2}$yr$^{-1/2}$] as a function of $\p$~[eV] (bottom) and frequency $\nu$~[Hz] (upper $x$ axis). Each bound is color-coded and labeled: UV/X-ray background (brown), $\gamma$-ray background (red), $y$-parameter (blue), $\mu$-parameter (black), Deuterium photofission (green), $^4$He photofission (magenta). See the text for more details.} \label{fig.260227.1}}
\end{figure}

\vspace{0.1cm}
{\it Bounds for high frequency at low redshift ($\p\gtrsim 10$\,{\rm eV}, $z<z_{\rm dec}$).}
From Eq.~\eqref{260122.3} we compute the flux of photons per unit energy and steradian (SM~\ref{sec.smbounds}). 
For the phenomenology of interest, we only consider sources at matter domination. Comparing  with the isotropic specific intensity $J_\nu$ resulting from the UV/X-ray background  synthesis model of Ref.~\cite{Faucher-Giguere:2019kbp}, 
\begin{align} \label{260227.1}
\left(\frac{E_{\text{rel}}}{M_\odot }\right)^{\!\!2}\!\Gamma\,\ep^2\!<\!\frac{175 \pi^2 H_0^2\Omega_mJ_{\nu}(E)}{3\, G^2 M_\odot^2 E^3\left(1-(E/\p)^{7/2}\right)}\,,
\end{align}
for every $\p$ and $E$  within its range $\p\geq  E\geq \ks/(1+z_{\text{a}})$. The absolute bound 
is obtained by minimizing the 
right-hand side with respect to $E$, cf. the brown line in Fig.~\ref{fig.260227.1}, labeled UV/X-ray background. 

At higher energies, the Fermi-LAT collaboration \cite{Fermi-LAT:2014ryh} measured the isotropic $\gamma$-ray background (IGRB) over the energy range from 100~MeV to 820~GeV. The IGRB consists of extragalactic emission too faint or diffuse to be resolved in current surveys, together with a residual approximately isotropic Galactic foreground. The relevant observable is the specific particle-number intensity, which can be obtained analogously to the specific intensity discussed above; the comparison yields the red curve in the right end of Fig.~\ref{fig.260227.1}, labeled $\gamma$-ray.

Energetic photons radiated by the GWs create electromagnetic showers that may photodisintegrate the light nuclei produced during BBN, constraining the high-energy GW parameters through the observed abundances of D, $^3$He, and $^4$He. 
Using the cascade spectrum and shower analysis of Refs.~\cite{Ellis:1990nb,Sarkar:1995dd}, we obtain the green (D) and magenta ($^4$He) bounds in Fig.~\ref{fig.260227.1}, labeled as D and $^4$He photofissions, respectively (SM \ref{app.260724.1}).

\vspace{0.1cm}
\noindent {\it Bounds from the microwave and radio backgrounds.} (SM~\ref{sec.smbounds})    We consider injection of photons happening after $z\sim3\times 10^6$, as before they simply thermalize. In the $\mu$-regime, $10^5\lesssim z\lesssim z_a\approx 3\times10^6$, the injected photons generate a chemical potential $\mu$, as Compton scattering is still efficient.  Making use of our previous formulas for the energy $\Delta\rho_\gamma$ and number of photons $\Delta n_\gamma$  injected during this interval, 
one can easily compute $\mu$  by imposing energy and photon number conservation \cite{Masso:1997ru,Chluba:2015hma}. Following  Ref.~\cite{Chluba:2015hma}, we also account for possible absorption of photons, 
and a typically small decrease in the resulting $\mu$ up to $z=0$  (SM~\ref{sec.smbounds}). From the COBE/FIRAS measurement~\cite{Fixsen:1996nj} one finds $|\mu|<9\times 10^{-5}\,,$ which implies the bound  shown by the black line in Fig.~\ref{fig.260227.1}.

In the range $10^3\lesssim z\lesssim z_a\approx 10^5$, Compton scattering can not thermalize the electrons with the injected photons. Typically, these photons transfer their energy to the plasma, which is then redistributed (Comptonized) among the photons in the background, giving rise to a  $y$-distortion, which COBE/FIRAS constrained to $|y|<1.5\times10^{-5}$~\cite{Fixsen:1996nj}. Adapting the results of Refs.~\cite{Peebles:1994xt,Chluba:2015hma} to our \emph{continuous radiation} of photons from the periodic and homogeneous distribution of GWs, one can compute the $y$-parameter for injection in the range of interest.
The resulting bound is shown by the blue line in Fig.~\ref{fig.260227.1}, labeled CMB $y$-parameter.

%%%%%%%%%%%%%%%%%%%%%%%%%%%%%%%%%%%%%%%%%%%%%%%%%%%%%%%%%%%%%%%%%%%%%%%%%%%%%%%%%%%%%%%%%%%%%%%%%%%%%%
\vspace{.1cm}
\noindent
{\bf Discussion and outlook.} Gravitational waves, described as coherent graviton states within the EFT of gravity, decay in vacuum into pairs of photons and light fermions and scalars (if they exist). This effect vanishes in the classical limit, but can be recovered semiclassically (a calculation that was missing in the literature).  We have found that, even if compact binaries provide a realistic astrophysical setting for this phenomenon, the resulting photon luminosities are very weak, e.g., comparable to the Hawking luminosity of a black hole with the total mass that of the binary. We have also shown that GWs may produce ULDM at higher rates, thanks to the large occupation numbers available in these scenarios. The luminosities may correspond to stellar ones in the cases of large DM densities, as may be the case in the solitons of galactic centers~\cite{Bar:2018acw}. For generic populations of recurrent GW sources, we showed that they are constrained by the bounds shown in Fig.~\ref{fig.260227.1}.

It would be interesting to extend this analysis to GW depletion by other stochastic backgrounds and macroscopic quantum states, including time-dependent rate $\Gamma$ of GW sources distributed in space or strong-lensing amplification. We also hope these estimates prove useful for other high-energy classical GW sources (e.g.\ cosmic-string bursts) or scenarios that enhance the effect, such as large extra dimensions where gravity may become stronger at the $\mu$m scale ($\sim10^{15}\,\mathrm{Hz}$). The extension of the formalism beyond the limit of narrow bandwidth is feasible and left for future work. Finally, one could investigate fermion injection from an SGWB at different cosmic epochs, extending works such as \cite{Kolb:2023ydq,Redi:2026nzi}. 

To close, let us remark that, even if our first estimates are not particularly optimistic, detecting them could open a new avenue for GW detection, while the process becomes a new possibility to generate particles cosmologically. This justifies an exploration of all their phenomenology, which we hope will happen in the near future. 

%%%%%%%%%%%%%%%%%%%%%%%%%%%%%%%%%%%%%%%%%%%%%%%%%%%%%%%%%%%%%%%%%%%%%%%%%%%%%%%%%%%%%%%%%%%%%%%
\section*{Acknowledgements}
We would like to thank P.~A.~Cano for interesting discussions and J.~Garriga for comments on the semiclassical calculations. We would also like to thank N. Rodd for very valuable comments on a previous version of this work.
JAO would like to acknowledge partial financial support to the Grant PID2022-136510NB-C32 funded
by \\MCIN/AEI/10.13039/501100011033/ and FEDER, UE, and to the EU Horizon 2020 research and innovation program, STRONG-2020 project, under grant agreement no. 824093. 
This publication is part of the R\&D\&i project PID2023-146686NB-C31 funded by MICIU/AEI/10.13039/501100011033/ and by ERDF/EU.
IFAE is partially funded by the CERCA program of the Generalitat de Catalunya.
This work is supported by ERC grant ERC-2024-SYG 101167211. Funded by the European Union. Views and opinions expressed are however those of the author(s) only and do not necessarily reflect those of the European Union or the European Research Council Executive Agency. Neither the European Union nor the granting authority can be held responsible for them.
D.B. acknowledges financial support from the Spanish Ministry of Science and Innovation (MICINN) through the Spanish State Research Agency, under Severo Ochoa Centres of Excellence Programme 2025-2029 (CEX2024001442-S).

%%%%%%%%%%%%%%%%%%%%%%%%%%%%%%%%%%%%%%%%%%%%%%%%%%%%%%%%%%%%%%%%%%%%%%%%%%%%%%%%%%%%%%%%%%%%%%%
\bibliographystyle{unsrturl}
\bibliography{references}

\onecolumngrid
\section*{Supplemental Material}
%\twocolumngrid
\setcounter{section}{0}
\setcounter{secnumdepth}{3}

\makeatletter
\renewcommand{\theHsection}{supp.\Alph{section}}
\renewcommand{\theHequation}{supp.\arabic{equation}}
\renewcommand{\theHfigure}{supp.\arabic{figure}}
\renewcommand{\theHtable}{supp.\arabic{table}}
\makeatother

\renewcommand{\theequation}{S\arabic{equation}}
\setcounter{equation}{0}
\setcounter{figure}{0}
\setcounter{table}{0}

    %%%%%%%%%%%%%%%%%%%%%%%%%%%%%%%%%%%%%%%%%%%%%%%%%%%%%%%%%%%%%%%%%%%%%%%%%%%%%%%%%%%%%%%%%%%%%%%
\subsection{Propagator of the coherent state}
\label{app.260319.1}

We show that the coherent-state mode does not propagate in diagram~2) of Fig.~\ref{fig.150715.1}, the only diagram with a graviton propagator connecting the two vertices $x$ and $y$. According to the standard expansion in perturbation theory for this diagram, we have two time orderings, times the product of the two graviton fields involved in the propagator. We develop the argument for a scalar field $\phi$ for simplicity.  See \cite{Ai:2025xla} for a recent application of in-medium effects related to the infrared structure of matrix decays in the presence of a GW background.

Discretizing in a volume $V$ with $a(\vk)=a_\vk\sqrt{2kV}$, the field reads
\begin{align}\label{260319.1}
\phi(x)=\sum_\vk\frac{1}{\sqrt{2kV}}\left(a_\vk e^{-ikx}+a_\vk^\dagger e^{ikx}\right).
\end{align}
In the coherent state $|f\rangle$, with $a_\vk|f\rangle=f_k|f\rangle$ and $f_k=f(k)/\sqrt{2kV}$, the time-ordered product of two fields splits into a free-mode part (proportional to $\delta_{\vk,\vk'}$) and coherent-mode terms bilinear in $f_k^{(*)}$:
\begin{align}\label{260319.2}
&\theta(x^0\!-\!y^0)\langle f|\phi(x)\phi(y)|f\rangle
=\theta(x^0\!-\!y^0)\sum_{\vk,\vk'}\frac{1}{2V\sqrt{kk'}}
\nn \\
&\times \Big( \nn e^{-ikx}e^{-ik'y}f_kf_{k'}+e^{ikx}e^{ik'y}f_k^*f_{k'}^*
+e^{ikx}e^{-ik'y}f_k^*f_{k'}+e^{-ikx}e^{ik'y}f_kf_{k'}^*
+\delta_{\vk,\vk'}e^{-ik(x-y)}\Big).
\end{align}
The expression for $\theta(y^0-x^0)$ is obtained by exchanging $x\leftrightarrow y$, and the dummy labels $\vk\leftrightarrow\vk'$ in the coherent terms. Then, the coherent-mode contributions are in fact identical for both orderings. The free-mode ($\delta_{\vk,\vk'}$) piece is the standard propagator already included in the vacuum amplitude of Eq.~\eqref{250530.3}~\cite{Skobelev:1975gpi} and is not considered further.

Spatial integrations at each vertex enforce three-momentum conservation, $\vK=\pm \vk$ and $\vQ=\pm \vk'$, so that $|\vK|=k$ and $|\vQ|=k'$, where $\vK$ ($\vQ$) is the total incoming (outgoing) three-momentum. Setting $K^0=k_1^0+k_2^0$ and $Q^0=q_1^0+q_2^0$, and performing the time integrations with the representation $\theta(x^0)=\frac{1}{2\pi i}\int dz\,e^{izx^0}/(z-i\epsilon)$, $\ep\to 0^+$$,$ we evaluate the coherent-mode contributions separately for each bilinear combination. For instance, consider  the $f_k^*f_{k'}^{}$ terms. Denoting by $I_{1a}$ and $I_{2a}$ the contributions from $x^0>y^0$ and $y^0>x^0$, respectively, the integration over these time variables gives,
\begin{equation}
    \label{260319.5}
I_{1a}=-i2\pi\,\delta(Q^0\!-\!K^0\!+\!k\!-\!k')\,f_k^*f_{k'}\,\frac{1}{K^0-k-i\epsilon}\,, ~~~~~
I_{2a}=\;i2\pi\,\delta(Q^0\!-\!K^0\!+\!k\!-\!k')\,f_k^*f_{k'}\,\frac{1}{K^0-k+i\epsilon}\,,
\end{equation}
and their sum is
\begin{align}\label{260319.5b}
I_{1a}+I_{2a}=4\pi^2\,f_k^*f_{k'}\,\delta(Q^0\!-\!K^0\!+\!k\!-\!k')\,\delta(K^0\!-\!k)\,.
\end{align}
Here, the imaginary parts have canceled. Because of the last Dirac-delta function $K^0=k$, which combined with $|\vK|=k$  gives $s=K^{0\,2}-\vK^2=0$, i.e.\ the intermediate graviton is forced on shell. Moreover, $K^0=|\vK|$ requires the two incoming momenta to be collinear, implying helicity conservation, which cannot be satisfied because $\pm2\pm2\neq\pm2$. The amplitude also vanishes at $s=0$, due to the derivative nature of gravitational couplings [cf.\ Eq.~\eqref{250530.3}].

The remaining three bilinear combinations ($f_kf_{k'}^*$, $f_kf_{k'}$, $f_k^*f_{k'}^*$) are handled analogously. In each case, the sum over both time orderings produces a $\delta$-function that forces the intermediate line on shell: either $K^0=k$ (as above) or $K^0=-k$, which contributes only at threshold $K^0=k=0$ where the amplitude again vanishes by the same derivative-coupling and helicity arguments. Therefore, the coherent-state mode does not propagate in diagram~2) of Fig.~\ref{fig.150715.1}.

The aforementioned cancellation originates from the displacement properties of $\hat D(f)$, which imply that $\langle f|T \hat h (x)\hat h (y) |f\rangle=h_{\rm c}(x) h_{\rm c}(y) +\langle 0|T \hat h(x) \hat h(y) |0\rangle$. Hence, the classical contribution in the propagator of 2) in Fig.~\ref{fig.150715.1} factorizes into two on-shell fields (one-point functions).  Momentum conservation at the three-graviton vertex forces $K$ carried by this intermediate coherent mode to be null, 
$K^2=s=0$, which requires the incoming gravitons to be collinear — precisely where the three-graviton vertex vanishes.

{\bf Corollary:} As noted above, each cancellation occurs independently for every combination $f_k^{(*)}f_{k'}^{(*)}$. This allows us to arrive at the same conclusion about the cancellation of the coherent-state contribution in the graviton propagator even if the initial and final coherent states were different. This is important when considering the total probability for $\gamma\gamma \,f \to gg$, and also in our results for the SGWB.

%%%%%%%%%%%%%%%%%%%%%%%%%%%%%%%%%%%%%%%%%%%%%%%%%%%%%%%%%%%%%%%%%%%%%%%%%%%%%%%%%%%%%%%%%%%%%%%
\subsection{Integrations to calculate $dP/dt$,  Eq.~\eqref{250717.10}}
\label{sec.260611.1} 
When an annihilation operator $\hat a_\lambda(\vk)$ present in the Fourier decomposition of the graviton field acts on $|f\ra$, it gives rise to
\begin{align} \label{250715.3}
\int\frac{{\di}^3k}{(2\pi)^32k}f^{s}(k)\ve^\lambda_{\mu\nu}(\vk)e^{-ik\cdot x}|f\ra\,,
\end{align}
which is a building block of the fusion amplitude in Eq.~\eqref{250715.7}. To compute the decay probability, we introduce the total and relative momenta for each pair of gravitons, 
\begin{align} \label{250716.3}
\vK^{(\smash{\prime})}=\vk^{(\smash{\prime})}_1+\vk^{(\smash{\prime})}_2~, ~\vk^{(\smash{\prime})}=\frac{1}{2}(\vk^{(\smash{\prime})}_2-\vk^{(\smash{\prime})}_1),
\end{align}
and momentum conservation implies $\vK'=\vK$. The Lorentz-invariant phase space of the two photons ${\di}Q_\gamma$ in the  CM reads 
$\displaystyle{
{\di}Q_\gamma=\frac{{\di}\Omega_\gamma}{32\pi^2}=\frac{\sin\theta_\gamma {\di}\theta_\gamma {\di}\phi_\gamma }{32\pi^2}}\,,
$
where $\theta_\gamma$ and $\phi_\gamma$ are the polar and azimuthal angles of the momentum of the first photon in the CM frame. Then, from Eq.~\eqref{250715.7}  
 we have
%\begin{widetext}
\begin{align} \label{250716.4}
P&\!=\!\!\frac{G^2}{16384\pi^9\ks^4}\!\!\int \!\! d^3K\!\!\!\int \! \!d^3k\!\!\int \!\! d^3k' f_{\p}\!(k_1)f_{\p}\!(k_2)f_{\p}\!(k'_1)f_{\p}\!(k'_2)(2\pi)\delta(k^0_1+k^0_2-k'^{0}_1-k'^{0}_2)\!\! \int \!\! d\Omega_\gamma  \frac{({\rm s}^2-2{\rm t}{\rm u})(\bar{{\rm s}}^2-2\bar{{\rm t}}\bar{{\rm u}})}{{\rm s}\bar{{\rm s}}},
\end{align}
%\end{widetext}
where  $\bar{{\rm s}}=(k'_1+k'_2)^2$, $\bar{{\rm t}}=(k'_1-q_1)^2$,  $\bar{{\rm u}}=-\bar{s}-\bar{t}$, and we have set 
${|\vk_i|=|\vk'_i|=\ks}$ in the denominators. With $\Dk/\ks\ll1$, the remaining energy-conserving delta function is treated as $2\pi\delta(0)\to T$, the interaction time, so that $P/T$ is the transition probability per unit time. 
This approximation improves for narrower bands since $T\sim1/\Dk$, provided $1/\Dk\gg\tau$, the reaction time, c.f. Eq.~\eqref{250717.11b}. 

With $\Dk/\ks\ll1$ it follows that 
$s=\bar{s}=4\ks^2-\vK^2\,,~\vK\cdot \vk=\vK\cdot\vk'=0\,.
$
The photon phase-space integration in $\Omega_\gamma$ in Eq.~\eqref{250716.4} is performed in the CM frame.  
The integrations over $|\vk|,|\vk'|$ and the polar angles are restricted by the narrow distributions $f^s(k)$. Writing $|\vk_i|=\ks+\delta k_i$ and $|\vk'_i|=\ks+\delta k'_i$,  with $|\delta k_i|,|\delta k'_i|\le\Dk/2$, 
the integrations over $\delta k$ and $\delta\theta$ are  straightforward. 
An analogous result holds for the primed variables. Finally, one has to do the integration over $\vK$ in the range $0\le|\vK|\le2\ks$  for which $s=4\omega_s^2-\vK^2$ has to be employed.  
Having done the integrations in Eq.~\eqref{250716.4}, $dP/dt$ given in Eq.~\eqref{250717.10} results.

By solving the differential equation for $N(t)$ from Eq.~\eqref{250717.10}, the radiated photon number is
\begin{align} \label{250717.11full}
N_\gamma(t)=\frac{G^2N_0^2\,6\ks^3\Dk^2 t/25\pi}{1+\frac{6G^2N_0\ks^3\Dk^2}{25\pi}t }\,,
\end{align}
whose first-order expansion gives Eq.~\eqref{250717.11} of the main text.

For $E_{\rm rel}=M_\odot$ and $\epsilon=0.15$ \cite{LIGOScientific:2016aoc}, the depleting time $\tau_h=\tau E_{\rm rel}/\omega_s$ is shorter than the age of the universe for $\omega_s\gtrsim10^4\,{\rm eV}$. In the linear approximation of Eq.~\eqref{250717.11},  $N_\gamma(\tau_h)=N_0$.

%%%%%%%%%%%%%%%%%%%%%%%%%%%%%%%%%%%%%%%%%%%%%%%%%%%%%%%%%%%%%%%%%%%%%%%%%%%%%%%%%%%%%%%%%%%%%%%
\subsection{SGWB depletion including cosmological expansion}
\label{sec.smsgwb}

The SGWB density matrix in the Glauber--Sudarshan $P$-representation reads \cite{Glauber:1963tx,Sudarshan:1963ts} (see also \cite{Cheong:2024ose})
\begin{align}   \label{260401.1}
  \hat\rho &= \int\!\bigg[\prod_{\si,\vk} d^2\!\alpha_{\si\vk}\,\mathcal{P}(|\alpha_{\si\vk}|)\bigg]
         \ket{\{\alpha_{\si\vk}\}}\!\bra{\{\alpha_{\si\vk}\}}\,,
\end{align}
where $\mathcal{P}(|\alpha_{\si\vk}|) 
    = \frac{1}{\pi N_{\vk}}\exp\!\left(-\frac{|\alpha_{\si\vk}|^2}{N_{\vk}}\right)$, $\ket{\{\alpha_{\si\vk}\}}$ is a coherent multimode state with 
$a_{\si\vk}\ket{\{\alpha_{\si\vk}\}} = \alpha_{{\si\vk}}\ket{\{\alpha_{\si\vk}\}}$
(in the discretized case  $a_{\si\vk}=a_\si(\vk)/\sqrt{2kV}$), and
$N_{\vk}=\la \{\alpha_{\si\vk}\}|a^\dagger_{\si\vk}a_{\si\vk}|\{\alpha_{\si\vk}\}\ra$. 

For $N_k\equiv N_\vk\gg1$ we neglect the zero-point term.
Given the previous density matrix, one finds $\displaystyle{\frac{d\sgw}{d\ln\! k}= \frac{k^4}{\pi^2}\!\left(N_k+\frac{1}{2}\right)}$. The effects of the expansion of the Universe are easily absorbed in the variable $\rho^{\rm co}(z)$, and  $d\rho^{\rm co}/d\!\ln\! k=(1+z)^{-4}d\sgw/d\!\ln \!k$ and its integral  $\rho^{\rm co}=\int_{-\infty}^{+\infty}\! d\!\ln\! k\, d\rho^{\rm co}/d\!\ln\! k$, the depletion rate of Eq.~\eqref{260401.3} implies
\begin{align} \label{260401.4}
  \frac{d^2\rho^{\text{co}}}{dz\,d\!\ln\! k}
    = \frac{16\pi G^2}{15 H(z)}(1+z)^4\,k\,\rho^{\text{co}}\;
       \frac{d\rho^{\text{co}}}{d\!\ln\! k},
\end{align}
having an extra factor $\hbar\, c^{-7}$ in  SI units ($k=\hbar\omega$).  With $y(k,z)=d\rho^{\text{co}}(z)/dk$ and $r(z)=16\pi G^2(1+z)^4/15H(z)$, the  solution to Eq.~\eqref{260401.4} is 
$y(k,z) \!=\! y(k,0)\, e^{-k\,\Phi(z)}$, $
\Phi(z)\! =\! -\!\int_0^z \!dz'\, r(z')\,\rho^{\text{co}}(z')
$, so that $\rho(z)\equiv\rho(\Phi(z))$ is the Laplace transform of $y(k,0)$ at $\Phi(z)$. This exact (formal) solution can be expanded perturbatively to find the results in the main text.

%%%%%%%%%%%%%%%%%%%%%%%%%%%%%%%%%%%%%%%%%%%%%%%%%%%%%%%%%%%%%%%%%%%%%%%%%%%%%%%%%%%%%%%%%%%%%%%
\subsection{Radiation of scalars: helicity amplitudes and rate}
\label{sec.smscalars}

The elementary helicity amplitudes for $gg\to\phi\phi$ follow from the Gross--Jackiw graviton--scalar Compton calculation~\cite{Gross:1968in} crossed to the ${\rm t}$-channel,
\begin{align}
  \mathcal{A}^{{\rm t}}_{++} =
  -\frac{8\pi G\, m^{4}\, {\rm s}}{({\rm t}-m^{2})({\rm u}-m^{2})}=\mathcal{A}^{{\rm t}}_{--}, ~~~~~~~~
  \mathcal{A}^{{\rm t}}_{+-} =
  -\frac{8\pi G\,(m^{4}-{\rm t}{\rm u})^{2}}{{\rm s}\,({\rm t}-m^{2})({\rm u}-m^{2})}=\mathcal{A}^{{\rm t}}_{-+},
\end{align}
the latter equalities follow from parity symmetry.

    Modeling the GW as a narrow-band coherent state ($\Dk/\p\ll 1$) of profile $f^s(k)$, and the ULDM background via a Glauber--Sudarshan $P$-representation, including the inverse process $\phi\phi\to gg$ with $g\in|f\ra$, summing helicities, and integrating over  $|f\ra$ and the scalar phase space, the rate is
\begin{align}   
\frac{dP}{dt}= \frac{G^{2}N^2(\delta \p)^{2} (1+2N_\phi)}{15360\pi\p^2}&
\int_{4m^{2}}^{4\p^{2}}\!\frac{d {\rm s} }{{\rm s} ^{3/2}\sqrt{4\p^{2}-{\rm s} }}\biggl[\, {\rm s} ^{3/2}
\times\! \si({\rm s} )(912m^{4}\!-\!136m^{2}{\rm s} \!+\!7{\rm s} ^{2})\! \nn\\ & 
+\! 7680m^{7} \mathrm{arccot}\frac{2m}{\sqrt{{\rm s} -4m^{2}}}+ 480 m^{4}({\rm s} -14m^{2})\sqrt{{\rm s} }     \,    \mathrm{arcsinh}\sqrt{\frac{{\rm s} }{4m^{2}}-1}\biggr],
\end{align}
with $\si({\rm s} )\!=\!\sqrt{1\!-\!4m^2/{\rm s}}$. The band ($m\ll\p$) and resonant [$m\leq\p\lesssim m(1+v^2/2)$] regimes quoted in the main text follow from this expression.

%%%%%%%%%%%%%%%%%%%%%%%%%%%%%%%%%%%%%%%%%%%%%%%%%%%%%%%%%%%%%%%%%%%%%%%%%%%%%%%%%%%%%%%%%%%%%%%
\subsection{Radiation density from homogeneous, isotropic, periodically recurring GW distributions}
\label{sec.smbounds}

The total energy radiated by GW$_j$ and observed at redshift $z$ is $E_j(z)= W_\gamma(1+z)\int_{z}^{z_j}{\di}z' \frac{(1+z')^{4}}{(1+z_j)^6\,H(z')}$, contained within the GW propagation sphere of physical radius $R_j(z)= \frac{1}{1+z}\int_{z}^{z_j}\frac{{\di}z'}{H(z')}$. Summing over all sources causally connected with an observation point $(\vx,t)$, i.e.\ those satisfying $\int_{t_j}^t \frac{{\di}t'}{a(t')}\geq |\vx-\mathbf{r}|$, the total energy density is
\begin{align} \label{251030.5}
\Delta \rho_\gamma(z)&=\int {\di}^3 r\sum_{t_j\geq t_a}^t\tilde{n}\, \theta\!\left(\int_{t_j}^t \frac{{\di}t'}{a(t')}-|\vx-\mathbf{r}|\right)\frac{W_\gamma(1+z)}{\frac{4}{3}\pi R_j^3}\int_z^{z_j} {\di}z'\frac{(1+z')^4}{(1+z_j)^6\,H(z')}\,.
\end{align}
Writing $t_j=t_a+j/\varkappa$, with $\varkappa$ the frequency of  generation of GWs, replacing the sum by an integral (valid for $1/\varkappa(t-t_a)\ll 1$ and $H(z)/\varkappa \ll 1$), and performing the integration over $\mathbf{r}$, Eq.~\eqref{251030.5} can be written as Eq.~\eqref{260122.3}, where 
\begin{align}
\label{eq:Fzomega}
F(z,\omega)=&H_0^2\int_{z}^{z_a}\!dz''\frac{1+2(e^{\frac{\bar\omega_s}{(1+z'')}}-1)^{-1}}{(1+z'')^7H(z'')}  \int_{z}^{z''}\!dz' \frac{(1+z')^4}{H(z')},
\end{align}
with $z_a=z(t_a)$ and $\bar{\omega}_s=\omega_s/(k_BT_0)$.

The flux of photons per unit energy and steradian follows from Eq.~\eqref{260122.3},
\begin{align} \label{260113.2}
\frac{dF_E}{dE d\Omega}&=\frac{(1+z)^4\,W_\gamma \Gamma}{4\pi} \int_{z}^{z_a}\frac{dz''(1+2(e^{\frac{\bar\omega_s}{(1+z'')}}-1)^{-1})}{(1+z'')^7 H(z'')}
\int_{z}^{z''}\!\!\! dz'\frac{(1+z')^4}{H(z')}\,\delta\left(E-\p\frac{1+z}{1+z''}\right)\,.
\end{align}

Since our injection of photons is not instantaneous, but extended in time, it may be useful to explicitly describe how the $y$ and $\mu$ distortions were derived.  

To compute $\mu$ from the photons injected by GW radiation, we impose energy and photon-number conservation, linearizing with respect to the small perturbations $\delta T$ and $\mu$, see e.g. \cite{Masso:1997ru}. When these photons are of very low enough frequencies, double-Compton scattering (DC) and Bremsstrahlung (BR) are very efficient in absorbing photons before they can be reshuffled towards higher frequencies by Compton scattering. This is accounted for by a survival function $P_s(\bar{\omega}_s/(1+z''),z)$ responsible of injecting photons at larger energies~\cite{Chluba:2015hma}.
In addition, following~\cite{Chluba:2015hma}, we also take into account the fact that the amplitude of $\mu$ slowly reduces due to the non-particle-number conserving DC and BR emissions/absorptions up to its present value. This is accounted for by the distribution visibility function,  $J_{\text{bb}}(z)$, for a photon injected at redshift $z$,
\begin{align}\label{260628.2}
J_{\text{bb}}(z)&\approx 0.983 e^{-(z/z_\mu)^{5/2}}\left(1-0.0381(z/z_\mu)^{2.29}\right)\,,
\end{align}
with $z_\mu\approx 1.98\times 10^6$, the thermalization redshift.

%\begin{widetext}
Then,  $\Delta n_\gamma^{\mu}$ and $\Delta\rho_\gamma^{\mu}$ are given by 
\begin{align}\label{260628.1}
\Delta n_\gamma^{\mu}&=\frac{W_\gamma^0\Gamma(1+z)^3}{\ks}   \int_{z}^{z_a}\frac{dz''(1+2(e^{\bar{\omega}_s/(1+z'')}-1)^{-1})}{(1+z'')^6 H(z'')}\, \int_{z}^{z''} \frac{dz'(1+z')^4}{H(z')}P_s(\tfrac{\bar{\omega}_s}{1+z''},z')J_{\text{bb}}(z')\,
\end{align}
with the $P_s(x,z)$ function coming from~\cite{Chluba:2015hma}, and
\begin{align} \label{260629.1}
\Delta \rho_\gamma^{\mu}&=(1+z)^4\,W_\gamma\Gamma \int_{z}^{z_a}\frac{dz''(1+2(e^{\bar{\omega}_s/(1+z'')}-1)^{-1})}{(1+z'')^7 H(z'')}\, \int_{z}^{z''} \frac{dz'(1+z')^4}{H(z')}J_{\text{bb}}(z')\,.
\end{align}
The $z=0$ value of $\mu$ then reads 
\begin{align}\label{260629.2}
\mu&=\frac{3\pi^2(45\zeta(3)\Delta\rho_\gamma^{\mu}-2\pi^4 k_B T_\gamma \Delta n_\gamma^{\mu})}{2(k_B T_\gamma)^4(\pi^6-405\zeta(3)^2)}\,.
\end{align}
From this expression, we can find the bounds appearing in Fig.~\ref{fig.260227.1}
%\end{widetext}

For the $y$-distortion, the Kompaneets equation yields
\begin{align} \label{260116.6}  \frac{dT_\gamma}{dt}&=\frac{\sigma_t n_e k_B}{m_e}  T_\gamma(T_e-T_\gamma)-T_\gamma H\,,\\  \label{260116.7}  \frac{dT_e}{dt}&=-\frac{8 \sigma_t u_\gamma}{3 m_e}(T_e-T_\gamma)+\frac{2}{3n_e k_B}\frac{d\Delta \rho_\gamma}{dt}-2T_e H\,, \end{align} 
where $\sigma_t$ is the Thomson cross section, $m_e$ the electron mass, $n_e$ the electron number density, and $u_\gamma=\pi^2 k_B^4 T_\gamma^4/15$ the blackbody energy density. From Eq.~\eqref{260116.7} the electron cooling time is $t_{\rm cool}=3m_e/(8\sigma_t u_\gamma)\sim 10^{20}\,z^{-4}\,\text{s}\ll H(z)^{-1}$, while the analogous timescale for $T_\gamma$ in Eq.~\eqref{260116.6} is much longer since $n_e/n_\gamma\sim 10^{-9}$, and the GW photon-radiation decay time is $t_{\text{gw}}=\Delta\rho_\gamma/(d\Delta\rho_\gamma/dt)\sim H^{-1}$. Treating $t_\gamma$ and $t_{\text{gw}}$ adiabatically relative to $t_{\text{cool}}$ gives $\displaystyle{k_B\Delta(t)\approx \frac{m_e}{4\sigma_t n_e u_\gamma}\frac{d\Delta\rho_\gamma}{dt},}$ with $\Delta(t)=T_e-T_\gamma$. The spectral distortion parameter $y$ is then, neglecting 1 in front of $z'$ and integrating by parts,
\begin{align} \label{260118.4}
\!\!y(z,z_a)&\!=\!
\!-\frac{15\zeta(3)}{2\pi^4 n_\gamma^0 k_B T_\gamma^0}
\!\left(\!\!
\left.\frac{\Delta\rho_\gamma(z')}{{z'}^4}\right|_{z}^{z_a}\!\!\!\!+4\!\!\int_z^{z_a} \!\!\! dz'\frac{\Delta\rho_\gamma(z')}{{z'}^5}
\!\right)\!.
\end{align}

%\begin{widetext}
To provide more accurate results, we modify this expression following Ref.~\cite{Chluba:2015hma} to account for the energy transfer to the electron plasma by the injected photons, differentiating between low frequencies,  where free-free dominates, and higher frequencies, where the electron recoil is more important. Eventually, this amounts to the change
    $d\Delta \rho_\gamma/dz$, we have to change it in the following way:
\begin{align}\label{260630.2}
\frac{d\Delta\rho_\gamma}{dz}&\rightarrow\frac{d\Delta\rho_\gamma^{\text{y}}}{dz}= \frac{4\Delta\rho_\gamma^{\text{y}}}{1+z}-\frac{W_\gamma\Gamma (1+z)^8}{H(z)}\int_z^{z_a}\frac{dz''(1+2(e^{\bar{\omega}_s/(1+z'')}-1)^{-1})}{(1+z'')^7H(z'')}h\left(\frac{\bar{\omega}_s}{1+z''},z''\right)\,,
\end{align}
where $h(x,y)$ is given in~\cite{Chluba:2015hma} and with
\begin{align}
\Delta\rho_\gamma^{\text{y}}&=(1+z)^4W_\gamma^0\Gamma\int_z^{z_c}dz''\frac{1+2\left(e^{\bar{\omega}_s/(1+z'')}-1\right)^{-1}}{(1+z'')^7H(z'')}\int_z^{z''}dz'\frac{(1+z')^4}{H(z')}h\left(\frac{\bar{\omega}_s}{1+z''},z'\right)\,.\nn
\end{align}
%\end{widetext}
The resulting improvement of including this more accurate description is generally modest, with the largest deviations occurring in the intermediate transition region, where $x\sim{\cal O}(1)$.

%%%%%%%%%%%%%%%%%%%%%%%%%%%%%%%%%%%%%%%%%%%%%%%%%%%%%%%%%%%%%%%%%%%%%%%%%%%%%%%%%%%%%%%%%%%%%%%
\vspace{2pt}
\subsection{Photofission of D and $^4$He.}
%\noindent\textit{Photofission of D and $^4$He.}\;
\label{app.260724.1}

The relevant thresholds for the photofission reactions of interest are
\begin{align} \label{260219.1}
&Q_{\gamma \rm{D}\to \rm p \rm n}=2.23~\text{MeV}~,~
Q_{\gamma\, ^4\rm He\to \rm p T}=19.8~\text{MeV}~,\\
&Q_{\gamma\, ^4\rm He\to \rm n \,^3\rm He}=20.6~\text{MeV}~,~Q_{\gamma \,^4\rm He\to \rm p n D}=26.1~\text{MeV}\,.\nn
\end{align}
We abbreviate $Q_D= Q_{\gamma\,\rm{D}\to\rm pn}$, $Q_1=20~\text{MeV}\approx Q_{\gamma\, ^4\rm He\to pT}\approx Q_{\gamma \,^4\rm He\to n\,^3\rm He}$, and $Q_2=Q_{\gamma\,^4\rm He\to\rm p n D}$.
Energetic photons generate electromagnetic cascades~\cite{Sarkar:1995dd,Ellis:1984er,Ellis:1990nb}: primary photons lose energy through $e^+e^-$ pair production on background photons, and the resulting pairs cool via inverse-Compton scattering, until photon energies fall below $
E_{\text{max}}\simeq m_e^2/(22\, T_\gamma(z))\,.$
Equating $E_{\text{max}}$ to $Q_D$, $Q_1$ and $Q_2$ defines $z_{\text{D}}=2.3\times 10^7$, $z_1=2.5\times 10^6$, $z_2=1.9\times 10^6$.

Using the cascade spectrum from Refs.~\cite{Ellis:1990nb,Sarkar:1995dd} and proceeding analogously to Eq.~\eqref{260113.2}, the photon spectrum emitted by GWs under radiation dominance is
\begin{align} \label{260219.5}
\frac{{\di}\Delta n_\gamma}{{\di}E}
&=\frac{W_\gamma\,\Gamma}{3\ks^2(1+z)\,H_0^2\Omega_r}\left(\frac{E}{\ks}\right)^3\left[1-\left(\frac{E}{\ks}\right)^3\right]\,.
\end{align}
For the cascade calculation, we use the comoving spectrum $d^*\Delta n_\gamma/dE=(1+z)^{-3}\,d\Delta n_\gamma/dE$.

For $z_{\text{D}}>z>z_1$ the dominant process is $\gamma \rm D\to \rm pn$. The abundance evolution in a comoving volume is~\cite{Sarkar:1995dd,Ellis:1984er,Ellis:1990nb,Lindley:1984bg}
\begin{align} \label{260219.12}
\frac{dX(\text{D})}{dt}&=-\frac{X(\text{D})}{n_e}\int_{E_{\text{max}}(z)}^{\ks} \!\! dE\,\frac{d^*\Delta n_\gamma}{dt\,dE}\\
&\times \int_{Q_{\text{D}}}^{E_\text{max}(z)} \!\!\! dE_\gamma \frac{dn_E}{dE_\gamma}\frac{\sigma_{\gamma\,\text{D}\to\text{pn}}(E_\gamma)}{\sigma_C(E_\gamma)}\,,\nn
\end{align}
with $\sigma_C$ the Compton cross section, photofission cross sections from Ref.~\cite{Laget:1978np}, and $n_e=\eta n_\gamma^0(1-Y/2)$ where $n^0_\gamma=2\zeta(3)(k_BT_0)^3/\pi^2$, $\eta=6.1\cdot10^{-10}$, and $Y\approx0.25$~\cite{Planck2018VI}. Integrating Eq.~\eqref{260219.12} to first order in $\delta_D$ and requiring $\delta_{\text{D}}<0.03$ yields the bound shown by the green line in Fig.~\ref{fig.260227.1}, labeled D photofission.

For $z<z_1$, photodisintegration of $^4$He into D and $^3$He dominates. Its abundance evolution can be derived similarly to Eq.~\eqref{260219.12}, with photodisintegration cross sections in vacuum from Refs.~\cite{Quaglioni:2003ym,gari_hebach_1981,Doran:1993oqt}.

%\end{widetext}

\end{document}